\documentclass{ws-procs9x6}

\begin{document}

\title{COSMIC RAY ACCELERATION IN SUPERNOVA REMNANTS}

\author{P. BLASI}

\address{INAF/Osservatorio Astrofisico di Arcetri\\
Largo E. Fermi, 5 Firenze, 50126 Italy\\
$^*$E-mail: blasi@arcetri.astro.it}

\begin{abstract}
We review the main observational and theoretical facts about acceleration of Galactic cosmic rays in supernova remnants, discussing the arguments in favor and against a connection between cosmic rays and supernova remnants, the so-called supernova remnant paradigm for the origin of Galactic cosmic rays. Recent developments in the modeling of the mechanism of diffusive shock acceleration are discussed, with emphasis on the role of 1) magnetic field amplification, 2) acceleration of nuclei heavier than hydrogen, 3) presence of neutrals in the circumstellar environment. The status of the supernova-cosmic ray connection in the time of Fermi-LAT and Cherenkov telescopes is also discussed.
\end{abstract}


\bodymatter

\section{Introduction}

The possibility that the bulk of cosmic rays (CRs) observed at Earth may be generated in supernova remnants (SNRs) dates back to the '30s \cite{Baade:1934p886} and with time it has acquired the rank of a paradigm, mainly because of the fact that on energetic grounds \cite{Ginzburg:1964p929} supernovae are the only class of sources in the Galaxy that can provide enough energy as to explain the CR flux observed at Earth. The basic requirements for the SNR paradigm are: 1) that SNRs may accelerate with typical efficiency of $\sim 10-20\%$; 2) that the spectrum of individual elements and the consequent all-particle spectrum are reproduced, including the presence of a knee at $\sim 3\times 10^{15}$ eV; 3) that the chemical abundances of nuclei are well described; 4) that the multifrequency observations of individual remnants, from the radio to the gamma ray band, are well described; 4) that the anisotropy induced by the spatial distribution of SNRs in the Galaxy is compatible with observations. 

The acceleration mechanism that is usually assumed to work in SNRs is diffusive shock acceleration (DSA) \cite{Bell:1978p1344,Blandford:1987p881}, but the energetic requirement that at least $\sim 10-20\%$ of the kinetic energy of the supernova shell is converted to CRs leads to realize immediately that the standard test-particle version of the theory describing this process is not applicable to the description of CR acceleration: the reaction of accelerated particles onto the accelerator cannot be neglected and in fact it is responsible for spectral features (such as spectral concavity) that may represent potential signatures of CR acceleration. There is another, possibly more important reason why DSA must include the reaction of accelerated particles: the standard diffusion coefficient typical of the interstellar medium (ISM) only leads to maximum energies of CRs in the range of $\sim$ GeV, rather than $\sim 10^{6}$ GeV (around the knee) required by observations. The only way that the mechanism can play a role for CR acceleration is if the accelerated particles generate the magnetic field structure on which they may scatter \cite{Lagage:1983p1348}, thereby reducing the acceleration time and reach larger values of the maximum energy. A non linear version of DSA including the dynamical reaction of accelerated particles on the shock was developed by many authors (see Ref. \refcite{Malkov:2001p765} for a review) and more recently completed with the inclusion of self-generation of magnetic field \cite{Amato:2005p112,Amato:2006p139,Caprioli:2008p123} and acceleration of nuclei other than Hydrogen \cite{Caprioli:2010p789}.

From the observational point of view the detection of narrow rims in the X-ray emission of several SNRs (Ref. \refcite{Parizot:2006p933} and references therein) has provided an important support to the idea that CRs may amplify the magnetic field close to the shock surface, thereby leading to reach higher energies: these rims are in fact most easily interpreted as the result of the synchrotron emission on a time scale comparable with the loss length of the highest energy electrons \cite{Volk:2005p968}. A simple estimate leads to magnetic fields of order $\sim 100-1000 \mu G$ downstream of the shock, which are hard to interpret as the result of solely the compression of the field at the shock surface since the ISM magnetic field is typically $\sim 1-10 \mu G$ and the compression at a strong shock is a factor $\sim 4$. It is important to realize that the CR induced magnetic field amplification occurs upstream of the shock; the perpendicular components of the field are further compressed at the shock surface. Moreover the overall structure of the rims, at least in the case of SN1006 appears to be inconsistent with the absence of magnetic field amplification upstream of the shock \cite{Morlino:2010p168}. However alternative explanations, involving acceleration at perpendicular shocks and magnetic field amplification due to turbulent eddies downstream of the shock \cite{Giacalone:2007p962} may still beb plausible explanations of the data. 

The rigidity dependent nature of DSA leads to predict higher energies for accelerated nuclei (if they get fully ionized), so that the knee may result from the superposition of the spectra of accelerated nuclei, if the magnetic field is amplified to sufficiently high levels. Unfortunately the problem of injection, hard enough for protons and electrons, becomes even harder for nuclei especially for those that may result from sputtering of dust grains \cite{Ellison:1997p609}. Some phenomenological attempts to calculating the contribution of nuclei in the context of non linear theory of DSA to the all-particle spectrum observed at Earth have recently been carried out \cite{Berezhko:2007p1010,Ptuskin:2010p1025,Caprioli:2010p789}. 

A crucial step towards confirming or rejecting the SNR paradigm might be made through gamma ray observations both in the TeV energy range, by using Cherenkov telescopes, and in the GeV energy range accessible to the Fermi gamma ray telescope. Gamma radiation can be produced mainly as a result of inverse Compton scattering (ICS) of relativistic electrons on the photon background and in inelastic proton-proton scatterings with production and decay of neutral pions. The present observational situation is rather puzzling and deserves some special discussion (see \S \ref{sec:gamma}): most gamma ray spectra observed by Fermi (see \cite{funk,tanaka} for reviews) (with some important exceptions, e.g. RX J1713-3946) hint to rather steep spectra of accelerated particles ($\propto E^{-\gamma}$, with $\gamma\sim 2.4-3$), which are not easy to accomodate in the context of NLDSA that predicts flat spectra, possibly even flatter than $E^{-2}$ at high enough energy \cite{Caprioli:2009p145}. Flat injection spectra would also lead to require a steep dependence of the Galactic diffusion coefficient on energy, which in turn is known to result in exceedingly large anisotropy \cite{Ptuskin:2006p620}. The most likely explanation for this discrepancy might lie in a rather subtle detail of the DSA theory, namely that the velocity relevant for particle acceleration is the velocity of waves with respect to the plasma. Usually the wave speed is negligible compared with the plasma velocity in the shock frame, but in the presence of magnetic field amplification this condition might be weakly violated. This is very bad news in that the spectral changes induced by this effect depend not only on the wave speed but on the wave polarization as well. In \cite{Caprioli:2010p789,Ptuskin:2010p1025} the authors show that there are situations in which the spectral steepening can indeed be sufficient to explain the observed spectrum of CRs and required by Fermi data on some SNRs. 

\section{Efficient CR acceleration in SNRs}

We first summarize the energetic requirements imposed by the SNR paradigm together with the diffusive propagation of CRs in the Galaxy. In a simple model in which the sources are all located in an infinitely thin disc with radius $R_{d}$ and the diffusion coefficient is constant within a halo of size $H$, the density of CRs is easily shown to be 
\begin{equation}
n_{CR}(E) \approx \frac{N(E){\cal R}_{SN}}{2 \pi R_{d}^{2}} \frac{H}{D(E)},
\label{eq:ncr}
\end{equation}
where ${\cal R}_{SN}$ is the rate of supernova explosions in the Galaxy, assumed to be spatially constant, $N(E)\propto E^{-\gamma}$ is the spectrum of CRs produced by an individual SNR (assumed to occur instantaneously)  and $D(E)$ is the diffusion coefficient, for which we adopt a typical functional form $D(E)=10^{28} D_{28} (\rho/3GV)^{\delta} cm^{2}/s$, where $\rho=E/Z$ is particle rigidity. In the following we concentrate our attention on ultrarelativistic protons, therefore $Z=1$ and their speed is the speed of light, $c$. The flux observed at Earth is $\phi(E)=c n_{CR}(E)/(4\pi)$. In Refs. \refcite{DiBernardo:2009p1027,Strong:2001p1041} the authors find that in order to fit at the same time the B/C ratio and the antiproton data one has to use $D_{28}/H_{kpc}\sim 1$, therefore one can write the CR flux  as 
\begin{equation}
\phi_{CR}(E) \approx 2.4 E_{51} \xi_{CR} R_{d,15}^{-2}{\cal R}_{SN,30} (\gamma-2) 3^{\delta} E_{TeV}^{-2.73} TeV^{-1}m^{-2}s^{-1}sr^{-1},
\label{eq:phi}
\end{equation}
where $E_{51}=E_{SN}/10^{51}erg$, $\xi_{CR}$ is the CR acceleration efficiency, ${\cal R}_{SN,30}$ is the SN rate in units of one every 30 years and $R_{d,10}=R_{d}/15 kpc$. It is easy to see that in order to fit the observed proton spectrum, $8.7\times 10^{-2} E_{TeV}^{-2.73} TeV^{-1}m^{-2}s^{-1}sr^{-1}$ one has to assume an acceleration efficiency $\xi_{CR}\sim 7\%$ for $\delta=1/3$, $\xi_{CR}\sim 11\%$ for $\delta=0.54$ and $\xi_{CR}\sim 58\%$ for $\delta=0.7$ for the reference values of the parameters. These should be considered as lower limits to the required efficiencies. These numbers immediately stress the need for a theory of particle acceleration that takes into account the possible dynamical reaction of the accelerated particles on the accelerator. 

    \refcite{Bell:1978p1344} another reason was discussed for having a reaction of CRs on the acceleration process: streaming instability induced by the accelerated particles leads to magnetic field amplification upstream of the shock (where a spatial gradient exists). In the absence of damping and if only resonant streaming instability is excited, the strength of the amplified magnetic field can be estimated as a function of the initial field $B_{0}$ (by using an extrapolation of quasi-linear theory) as \cite{Amato:2006p139}:
\begin{equation}
\delta B \sim B_{0} \sqrt{2 M_{A} \xi_{CR}},
\end{equation}
where $M_{A}\gg 1$ is the Alfvenic Mach number. For typical values of $M_{A}\sim 1000$ one obtains $\delta B/B_{0}\sim 30$ and the field is further amplified by compression when the fluid element crosses the shock from upstream to downstream. Larger values of the amplification factor can, at least in principle, be reached if non-resonant streaming instabilities are at work \cite{Bell:2004p737}. It is however not clear which role the field can play in this case in terms of scattering particles up to the knee energy since the unstable waves are produced at wavelengths much shorter than the Larmor radius of the particles generating them. Non linear effects in wave evolution can lead however to transporting energy towards longer wavelengths, though it is not clear whether this process is fast enough to occur within the precursor and thereby lead to efficient particle scattering, essence of the acceleration process. 

These two processes, dynamical reaction of accelerated particles and magnetic field amplification, are the two most important ones to take into account when describing non-linear particle acceleration in SNR shocks. The main predictions of the theory of non-linear DSA are the following: 1) spectral concavity induced by the momentum dependence of the diffusion coefficient in the upstream precursor; 2) effects related to magnetic field amplification (X-ray rims downstream, spatial profile of the upstream X-ray brightness, maximum energy of accelerated particles, possibly accessible through the gamma ray range); 3) lower temperatures of the thermal plasma downstream, induced by the simple fact that less energy is left for plasma heating after particle acceleration. 

\section{Observational signatures of efficient CR acceleration}

In this section we briefly discuss some of the observational evidence pointing towards efficient CR acceleration in individual SNRs. It is worth keeping in mind that while the SNR paradigm for the origin of CRs is usually quoted without any specific reference to the type of supernovae involved, the particle acceleration process is strongly dependent upon the environment in which the supernova goes off. For instance we can expect that the blast wave of a SN-type Ia has a larger Mach number than that of a type II SN, mainly because of the lower temperature of the ordinary ISM compared with that of the material around a massive star with powerful winds. For the same reason we can expect the gas density to be lower for type-II SNe. While type-II SNe are more frequent than type Ia, it is probably more difficult to observe a gamma ray emission of hadronic origin from these objects because of the lower gas density. This short warning is meant to stress that the constraints obtained on the particle acceleration efficiency of a few SNRs are hard to generalize to the whole zoo of SNRs. 

\subsection{X-ray emission from rims and precursor}

Non-thermal X-ray radiation is produced in SNRs through synchrotron emission of high energy electrons in the magnetic field around the shock. The emission is dominated by the region downstream of the shock where the magnetic field is stronger and is cut off at a frequency that, in the case of Bohm diffusion, is independent of the strength of the magnetic field:
\begin{equation}
\nu_{max}\approx 0.2 ~ u_{8}^{2} ~ \rm keV ,
\end{equation}
where $u_{8}=u_{sh}/(10^{8}cm/s)$ is the shock velocity in units of $1000$ km/s. The maximum energy of accelerated electrons depends on the strength of the amplified magnetic field and can be estimated as
\begin{equation}
E_{max}\approx 10 ~ B_{100}^{-1/2} ~ u_{8}~ \rm TeV,
\end{equation}
where $B_{100}=B/100 \mu G$ is the magnetic field in units of $100 \mu G$.

The emission region has a spatial extent at $\nu\sim \nu_{max}$ which is determined by diffusion and can be written as
\begin{equation}
\Delta x \approx \sqrt{ D(E_{max}) \tau_{loss}(E_{max})} \approx 0.04 ~B_{100}^{-3/2}~ \rm pc.
\end{equation}
The typical thickness of the X-ray rims found through high resolution observations is of order $\sim 10^{-2}$ pc, thereby leading to predicting values of the magnetic field of $100-300 \mu G$ downstream of the shock (or $\sim 20-80 \mu G$ upstream of the shock, if the field is amplified by the streaming of accelerated particles and further compressed at the shock). 

\begin{figure}
\includegraphics{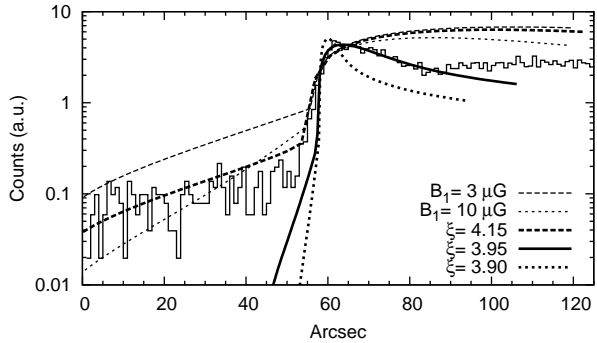}
\caption{Radial profile of the emission in the 0.8-2.0 keV band, extracted from a narrow strip around the rim of SN 1006 (thin black line). Overplotted are the theoretical predictions of NLDSA for 3 different injection efficiencies (thick lines) and those of test-particle theory for two different values of the pre-existing turbulent magnetic field upstream (thin lines).}
\label{fig:sn1006}
\end{figure}

In Fig. \ref{fig:sn1006} we show the X-ray brightness profile (histogram) of part of the rim of SN 1006 \cite{Katsuda:2009p1042}. The thick lines refer to the predictions of NLDSA \cite{Morlino:2010p168} for three values of the injection efficiency (larger values of $\xi$ correspond to lower efficiencies). The thin lines are the predicted radial profiles in test-particle theory for two values of the upstream magnetic field (as indicated). Two pieces of information arise from this figure: 1) the narrow rims ($\sim 10-20$ arcsec) downstream can only be reproduced for efficient particle acceleration scenarios; 2) the predicted  X-ray emission from the precursor region drops below the background for the same cases in which the rims are present. In other words, the non-detection of the precursor X-ray emission might be additional evidence for efficient acceleration. This is due to the fact that the spatial extent of the precursor is reduced when magnetic fields are amplified by CR streaming.

Although very interesting, this interpretation is not totally unique: it could be that the magnetic field is not amplified upstream by CRs, but rather enhanced due to fluid instabilities downstream in quasi-perpendicular shocks \cite{Giacalone:2007p962}. In this case the non-detection of the upstream emission could be due to a dominantly perpendicular topology of the magnetic field lines. 

\subsection{Collisionless shocks in partially ionized plasmas}

The blast waves produced by supernovae explosions give rise to collisionless shocks. The formation of the shock, as well as of the precursor is due to plasma processes which affect the ionized component of the plasmas involved. The neutral component, if present, is affected by the presence of the shock and by particle acceleration only through the process of charge exchange and, to a lesser extent, through ionization. A neutral atom can exchange an electron with an ion if the two are moving with different speeds. This can happen in three situations: 1) if the temperatures of ions and neutrals are different; 2) if ions and neutrals share the same temperature but two atoms lie on different parts of the thermal velocity distribution; 3) If the ions and neutrals have different bulk velocities.  

In the absence of CR induced precursor, the effect of charge exchange appears after shock crossing: downstream of the shock the ionized plasma is compressed and shock heated, while the neutral gas keeps moving with the upstream speed $u_{0}$ (neutrals might also have an intrinsic thermal motion, but for the present discussion we can neglect this point). At this point, charge exchange may take place in that cold neutral can lose an electron to a hot ion. As a result, a population of colder ions and one of warmer neutrals are produced. 

The observational signature of this phenomenon is clear in the so-called {\it Balmer dominated shocks}, where the Balmer lines are clearly detected. Charge exchange leads to the formation of a narrow line, corresponding to the neutrals that keep the same 'temperature' as they had upstream and did not suffer charge exchange, and a broad line corresponding to the neutrals that have been produced through charge exchange and were hot ions before the process. The measurement of the width of the narrow and broad Balmer lines allows one to measure the temperature of the two components. 

The situation becomes more interesting in the presence of cosmic rays: when the pressure of accelerated particles upstream of the shock generates the precursor, in the shock frame the bulk motion of ions is slowed down with respect to neutrals. Neutral atoms keep moving with the velocity that ions have at upstream infinity, $u_{0}$, therefore a difference in bulk velocity between the two species arises. At the shock crossing, the ionized plasma suffers shock heating and compression while the neutral component does not feel the shock and keeps moving with speed $u_{0}$. In the presence of CR acceleration, the temperature of the ions downstream is clearly lower that in the absence of CRs, simply because of energy conservation (there is now another component, CRs, into which the ram pressure $\rho_{0}u_{0}^{2}$ can be channelled). Therefore one can expect that the width of the broad Balmer line is somewhat smaller than in the absence of CRs. This phenomenon has been recently observed \cite{Helder:2009p660} in the SNR RCW86. 

Measurements of the proper motion of the shock lead to a shock velocity of $u_{0}=6000\pm 2800$ km/s, which in turn should imply a downstream temperature of $T_{2}=20-150$ keV or $T_{2}=12-90$ keV depending on whether protons and electrons reach or not thermal equilibrium downstream, and assuming that no CR acceleration is taking place. The temperature inferred from the width of the broad Balmer line is $T_{2}=2.3\pm 0.3$ keV. The authors interpreted this discrepancy as the result of effective CR acceleration at the SNR shock, with an estimated efficiency of $\sim 50-60\%$ (this numerical value is however rather model dependent). 

As pointed out above, in CR modified shocks some level of charge exchange is expected to take place in the precursor because of the difference in bulk velocity between the ionized plasma and the neutrals. The process is responsible for producing a population of warmer neutrals upstream. After shock crossing this leads to a narrow Balmer line which is broader than in the absence of a CR induced precursor. This effect was observed in several SNRs \cite{Sollerman:2003p615}. An independent signature of the same phenomenon was recently found in the Tycho SNR  \cite{Lee:2010p658}: the authors used the Hubble Space Telescope to measure the intensity of the broadened narrow Balmer line as a function of position around the shock and claim that there is substantial emission from the region in front of the shock. This finding could be either the signature of different bulk velocities between ions and neutrals (precursor) or of a different temperature of the two components. In this latter case, the ions could have been heated up due to the action of turbulent heating in the precursor. 

A quantitative investigation of these phenomena in the presence of realistic cosmic ray modified shock structures is under way (see Ref. \cite{Morlino:2010p1607} for preliminary results).
 
In the future the importance of the detection of anomalous widths of Balmer lines in SNR shocks will plausibly increase especially if a clear correlation with the detection of narrow non-thermal X-ray rims could be established. This correlation would represent a crucial confirmation that magnetic field amplification and efficient CR acceleration are related to each other.

\subsection{Maximum energy and the knee}

The spectrum of individual elements in CRs carries more information than the all-particle spectrum in terms of understanding particle acceleration in the sources. The spectrum of protons measured by KASCADE \cite{Apel:2009p1363} shows a pronounced decline at $\sim few\times 10^{6}$ GeV, which is likely to be produced by the acceleration process running out of steam. Although other experiments appear to disagree on the details of the measurement, they also show a decline in the same energy range. The situation for He nuclei is similar but it becomes rapidly worse for heavier nuclei. If the interpretation of this decline in terms of a maximum energy at the sources is correct, this implies that CR protons need be accelerated at least to $\sim few\times 10^{6}$ GeV in SNRs, in they are the sources. 

For Bohm diffusion, this result can be achieved in SNRs only if the magnetic field is amplified by $\sim 100$, a number surprisingly close to that inferred from the width of X-ray rims. This conclusion was also confirmed in the context of NLDSA \cite{Blasi:2007p144}. Since the maximum energy reached through diffusive shock acceleration scales with the charge of the nucleus, another consequence of this line of thought is that at energies above $\sim few\times 10^{6}$ GeV the chemical composition should become dominated by heavier nuclei and that the spectrum of Galactic CRs should eventually end at energies of order $\sim 10^{8}$ GeV with an iron dominated composition. At this point a transition to extragalactic CRs is expected. 

\section{Problematic aspects of the SNR paradigm}

Can we consider the SNR paradigm as proven right? As discussed in the previous section there are plenty of evidences that at least some of the SNRs we observe are accelerating CRs efficiently. What does not allow us to claim a full confirmation is the existence of several loose ends as well as the fact that other SNRs do not seem to be accelerating (hadronic) CRs to the level required by the paradigm. It is however worth stressing here again a crucial point. The simplicity of the SNR paradigm contrasts with the complexity of Nature in many different ways and while the general theory makes some very general predictions on a ``typical'' SNR, the behaviour of an individual SNR depends on many physical phenomena which are specific of the environment where the supernova explodes and are much harder to predict. Below we list some of the aspects that represent possible challenges for the SNR paradigm. 

\subsection{Gamma ray observations}\label{sec:gamma}

One of the observational signatures of SNRs as sources of CRs that has been long waited for is the detection of gamma radiation from the production and decay of neutral pions. Aside from the experimental difficulties in detecting the gamma rays from SNRs, it may be even more problematic to establish whether a detected flux can be unambiguously attributed to $\pi^{0}$ decay or rather to ICS of accelerated electrons. In the last few years the gamma ray detections of SNRs have proliferated both with Cherenkov telescopes \cite{Gallant:2010p1228,Carmona:2009p1279,Buckley:2010p1284} and with the Fermi-LAT telescope \cite{Funk:2010p1314}. 

The first case that was strongly suggestive of a hadronic origin of CRs was SNR RX J1713-3946, detected by HESS in the TeV energy region and later by Fermi \cite{funk} at lower energies. The combined spectrum appears to be relatively flat and probably of hadronic origin (see \S \ref{sec:xray} for more discussion on this point). Rather unexpectedly most of the SNRs detected by  Fermi have a rather steep gamma ray spectrum, with slope $\sim 2.4-3$, quite unlike the flat or even concave spectra predicted in the context of NLDSA. One might argue that the steep spectra are mostly observed in old SNRs with a nearby molecular cloud that serves as a target for inelastic CR collisions. This possibility is certainly to be kept in mind.

Whether such steep spectra may be compatible with having efficient CR acceleration and magnetic field amplification still remains to be investigated, but it is interesting to notice that, as we discuss in \S \ref{sec:spec}, the observed spectrum of CRs at Earth also leads to a preference of relatively steep injection spectrum. Flat injection implies in fact a too large CR anisotropy \cite{Ptuskin:2006p620}.

\subsection{X-ray lines}\label{sec:xray}

As mentioned above, gamma rays were detected from SNR RX J1713-3946. In Ref. \refcite{Morlino:2009p140} a detailed analysis of the arguments in favor and against a hadronic origin of this radiation was carried out. The authors conclude that a hadronic origin is favored but that there are some problems with it, the most important of which is that the predicted thermal X-ray bremsstrahlung emission exceeds observations unless the temperature of electrons is much smaller than that of protons. Indeed it can be expected that electrons are thermalized to a temperature $\sim m_{e}/m_{p}$ lower than for protons and that they gradually thermalize with protons on time scales as large as the time scale for Coulomb scattering, although other collisionless processes may lead to faster thermalization. In Ref. \refcite{Ellison:2010p636} it was argued that Coulomb scattering is sufficient to heat up the electrons to a temperature of $\sim 1$ keV during the expansion of the remnant, enough to excite emission lines whose intensity should have been detected. From the non-detection the authors conclude that the observed gamma rays cannot be of hadronic origin. 

\subsection{Spectra at Earth}\label{sec:spec}

In the simple case of a power law injection spectrum of CRs $\propto E^{-\gamma}$, the spectrum observed at Earth is $n(E)\sim E^{-\gamma-\delta}$, where the diffusion coefficient is assumed to be $D(E)\propto E^{\delta}$. This simple argument implies that the flatter the injection spectrum the larger must be the value of $\delta$. However $\delta> 0.5-0.6$ leads to excessive anisotropy of the CRs observed at Earth, thereby imposing a preference for relatively steep injection spectra. For instance for a Kolmogorov diffusion coefficient $D(E)\propto E^{1/3}$ the required injection spectrum must be $\sim E^{-2.4}$. As discussed above, such steep spectra are not easy to obtain in NLDSA with efficient acceleration and magnetic field amplification. The only possibility, recently suggested in Refs. \refcite{Ptuskin:2010p1025,Caprioli:2010p789} is that the magnetic field amplification also leads to introduce a relatively large velocity of the scattering centers responsible for particle diffusion in the shock region. This phenomenon leads to steeper spectra even in the context of NLDSA. The net effect is however to have lower magnetic fields and lower efficiencies of particle acceleration \cite{Caprioli:2010p789}. In Ref. \refcite{Caprioli:2010p789} the authors calculate the spectra of different nuclei at the source and after propagation: the injected spectra are steeper than in standard NLDSA, but they still require $D(E)\propto E^{0.54}$, in agreement with previous calculations \cite{Ptuskin:2010p1025} and sort of borderline from the point of view of anisotropy. 

It is important to keep in mind that calculations of the spectra of CRs released by SNRs into the ISM are highly uncertain because of the lack of a proper understanding of the processes responsible for the escape of CRs from the sources \cite{Caprioli:2009p145,Caprioli:2010p133,Drury:2010p1437}. 

\section{Conclusions}

We provided a short review of the main arguments in favor and against the so called SNR paradigm for the origin of Galactic CRs. There is a train of recent evidence that efficient CR acceleration takes place in several SNRs. Probably the most striking findings are the detection of narrow X-ray rims interpreted as evidence for magnetic field amplification, and numerous observational results on Balmer dominated shocks, which also lead to conclude that CRs are being accelerated effectively. 

On the other hand, there are a few pieces of the puzzle that do not appear to be in place: the non-detection of X-ray lines from SNR RX J1713-3946 is certainly problematic for a scenario in which the detected gamma rays are of hadronic origin. Fermi-LAT gamma ray observations of several SNRs with steep spectra also seems to be at odds with the standard predictions of NLDSA which is expected to describe efficient CR acceleration in SNR shocks. Steep spectra are also suggested by observations of CR anisotropy. We have discussed how the last two issues mentioned above could be understood by taking into account the finite velocity of the waves scattering particles close to the shock. 

\section*{Acknowledgments}
The author is grateful to his collaborators, E. Amato, D. Caprioli, G. Morlino, for precious scientific collaboration on the topics discussed here.

\bibliographystyle{ws-procs9x6}
\bibliography{ws-pro-blasi}

\end{document}